\documentclass[aps,prd,showpacs,nofootinbib,showkeys,superscriptaddress]{revtex4-2}
\usepackage{graphicx}
\usepackage{fixltx2e} 
\usepackage{definitions}
\usepackage{hyperref}
\usepackage{amsmath}
\usepackage{amssymb}
\usepackage{txfonts}
\usepackage{epstopdf}
\usepackage[utf8]{inputenc}
\begin{document}
\title{Equilibrium Chiral Magnetic Effect: spatial inhomogeneity, finite temperature,  interactions}
\author{Chitradip Banerjee}
\email{banerjee.chitradip@gmail.com}
\author{Meir Lewkowicz}
\author{Mikhail A. Zubkov}
\affiliation{Ariel University, Ariel 40700, Israel}

\date{\today}

\begin{abstract}
	We discuss equilibrium relativistic fermionic systems in lattice regularization, and extend the consideration of chiral magnetic effect
	to systems with spatial
	inhomogeneity and finite temperature. Besides, we take into account interactions due to exchange by gauge bosons. We find that the equilibrium chiral magnetic
	conductivity remains equal to zero.
	   
\end{abstract}

\maketitle

\section{Introduction}

The chiral magnetic effect (CME) \cite{Vilenkin,CME,Kharzeev:2013ffa,Kharzeev:2009pj,SonYamamoto2012} is one of the non - dissipative transport effects. Unlike most other members of this family it most likely does not appear in true equilibrium. Instead it appears, presumably, in a steady state out of equilibrium in the presence of both external electric field and external magnetic field \cite{Nielsen:1983rb}. Combination of these fields gives rise to chiral imbalance. The latter together with the magnetic field leads to electric current directed along the magnetic field. Experimental evidence of this effect is found in magnetoresistance of Dirac and Weyl semimetals, and its dependence on the angle between magnetic and electric fields \cite{ZrTe5}. The calculation of the CME conductivity \cite{Nielsen:1983rb,ZrTe5} demands reference to kinetic theory, and the chiral imbalance appears as a pure kinetic phenomenon. Yet it is not sufficiently clear, is this possible to describe this imbalance using the notion of chiral chemical potential. The resulting expression for the CME current is proportional both to the magnetic field squared  and to electric field. In the systems without external electric field the chiral magnetic effect may possibly be observed in the non - equilibrium systems with chiral chemical potential depending on time \cite{Wu:2016dam}.

The other non - dissipative transport effects have  been widely discussed in condensed matter physics and in high energy physics  \cite{Landsteiner:2012kd,semimetal_effects7,Gorbar:2015wya,Miransky:2015ava,Valgushev:2015pjn,Buividovich:2015ara,Buividovich:2015ara,Buividovich:2014dha,Buividovich:2013hza}. Some of them may also be observed  in the recently discovered Dirac and Weyl semimetals  \cite{semimetal_effects6,semimetal_effects10,semimetal_effects11,semimetal_effects12,semimetal_effects13,Zyuzin:2012tv,tewary}.
The experimental indications of the CME in relativistic heavy - ion collisions were discussed, for example, in \cite{Kharzeev:2015znc,Kharzeev:2009mf,Kharzeev:2013ffa}. As a fluctuation the CME has been reported using lattice simulations  \cite{Polikarp}.

As it was mentioned above, at the present moment it is widely believed that the equilibrium version of CME does not exist. This has been proven for the case of homogeneous systems at zero temperature (better to say, those systems that become homogeneous when external magnetic field is removed). In \cite{Valgushev:2015pjn,Buividovich:2015ara,Buividovich:2015ara,Buividovich:2014dha,Buividovich:2013hza} the proof has been given using lattice numerical simulations. In  \cite{Gorbar:2015wya} the same conclusion was obtained using analytical methods for the system of finite size with the specfic  boundary conditions in the direction of the magnetic field. In \cite{nogo} the absence of CME was reported for a certain model of Weyl semimetal. The contradiction of  equilibrium CME to the no - go Bloch theorem has been noticed in \cite{nogo2}. In \cite{Z2016} it has been shown that at zero temperature the CME current is a topological invariant. Its responce to any parameter (to the chiral chemical potential, as an example) is vanishing, which is an alternative proof. 

So far the question about the possibility to observe equilibrium CME current remains open for the systems at finite temperature with explicit dependence of lagrangian on coordinates. (The homogeneous systems at finite temperature have been considered in  \cite{Beneventano:2019qxm} using zeta regularization.) Moreover, the influence of interactions on the CME conductivity has not been considered in sufficient details. In the present paper we close this gap and demonstrate that the CME conductivity remains vanishing in equilibrium under these circumstances.  

Technically we rely on the machinery developed in \cite{ZW2019,ZZ2019,ZZ2019_}. 
Namely, we use Wigner transformed Green functions in order to express electric current and its response to external fields. Using this technique we will show that the response of total electric current to constant external magnetic field is a topological invariant even in the non - homogeneous systems, and even at finite temperature. Besides, we will demonstrate that interactions due to exchange by gauge bosons do not affect this conclusion at the one and two - loop level. The generalization of this consideration to the higher orders may be given along the lines of \cite{ZZ2019}.

Hystorically Wigner - Weyl calculus has been proposed in order to reformulate quantum mechanics using language of functions in phase space instead of the language of operators \cite{Weyl_1927}.
The so - called "deformation quantization" has been developed on the basis of this calculus (see \cite{Vassilevich_2008,Vassilevich_2015} and references therein).
One of the standard quantites of the Wigner - Weyl calculus is Wigner
function $W(q,p)$ that generalizes the notion of distribution
in phase space of classical mechanics \cite{Balazs_1984}.
It is worth mentioning that Wigner distribution can not be treated as probability distribution \cite{Zurek_1991}.
Nevertheless, it is possible to formulate the fluid analog of quantum entropy flux
	in phase space using Weyl-Wigner calculus \cite{Bernardini_2017}.
Von Neuman entropy and the other quantites of quantum information theory may be defined using Wigner - Weyl calculus  \cite{Bernardini_2019,Wlodarz_2003,Bernardini_2017}. We would also like to notice that various applications to quantum mechanics
\cite{Curtright_1998,Bastos_2008,Bernardini_2015,Bernardini_2017} have been followed by the applications of Wigner - Weyl calculus to anomalous transport in quantum field theory 
\cite{Lorce_2011,Lorce_2012,Buot_1990,Buot+Jensen_1990,Miransky+Shovkovy_2015,Prokhorov+Teryaev_2018}.

\section{Equilibrium inhomogeneous theory at zero temperature}

\subsection{Model setup}

We start from the fermionic lattice model with partition function written in momentum space \cite{Z2016}
\begin{eqnarray}
		Z &=& \int D\bar{\psi}D\psi \, {\rm exp}\Big(  \int_{\cal M} \frac{d^D {p}_1}{\sqrt{|{\cal M}|}} \int_{\cal M} \frac{d^D {p}_2}{\sqrt{|{\cal M}|}}\bar{\psi}^T({p}_1){\cal Q}(p_1,p_2)\psi({p}_2) \Big),\label{Z1}
\end{eqnarray}
By $|{\cal M}|$ we denote volume of four - dimensional momentum space $\cal M$. Both $p_1$ and $p_2$ are four - momenta. Here matrix elements of lattice Dirac operator $\hat{Q}$ are denoted by 
$$
\langle p_1| \hat{Q} | p_2 \rangle = {\cal Q}(p_1,p_2)
$$ 
It is worth mentioning that we use the relativistic units, in which both $\hbar $ and $c$ are equal to unity.  In addition, elementary charge $e$ is included to the definition of electric and magnetic fields. 

Wigner transformation of Green function ${\cal G}(p_1,p_2) = \langle p_1 | \hat{G} | p_2\rangle$, i.e. the Weyl symbol of operator $\hat{G}$ is defined as
\begin{equation} 
		{G}_W(x,p) \equiv \int_{\cal M} dq e^{ix q} {\cal G}({p+q/2}, {p-q/2})\label{GWx}
\end{equation}
In the similar way we define Weyl symbol of operator $\hat{Q}$:
$$
{Q}_W(x,p) \equiv \int_{\cal M} dq e^{ix q} {\cal Q}({p+q/2}, {p-q/2})$$
If the inhomogeneity is due to the slowly varying external gauge field $A(x)$ (when its variation at the distance of the order of lattice spacing may be neglected) we have
 ${Q}_W(x,p) = {Q}_W(p-A(x))\equiv {Q}(p-A(x))$. 
Under these conditions matrix element $Q(p_1,p_2)$ remains nonzero only when $|p_1-p_2|$ is much smaller than the size of compact momentum space. This property remains also for sufficiently weak  inhomogeneity caused by any other reasons. Then Wigner transformation of the Green function and Weyl symbol of Dirac operator obey the Groenewold equation \cite{ZS2020}
\begin{equation} 
		{G}_W(x,p) \star Q_W(x,p) = 1 \label{Geq}
\end{equation}
Here the Moyal product is introduced
\begin{equation}
		\star \equiv 
			e^{\frac{i}{2} \left( \overleftarrow{\partial}_{x}\overrightarrow{\partial_p}-\overleftarrow{\partial_p}\overrightarrow{\partial}_{x}\right )}
			\label{GQW}
\end{equation}
Let us notice associativity of the Moyal product, which allows to remove brackets in expressions containing the stars. Besides, we will use below the following property of the star product:
$$
\int d^D x d^D p {\rm Tr} A_W \star B_W = \int d^D x d^D p {\rm Tr} A_W  B_W
$$
This property is valid for any functions $A_W, B_W$ defined in phase space.

  We consider external electromagnetic potential that gives rise to magnetic field ${\bf{B}}({\bf{x}})=\nabla\times {\bf{A}}({\bf{x}})$ and to varying electric field ${\bf E}({\bf x}) = - \nabla \phi({\bf x})$. In Euclidean space $A_4 = - i \phi$, while spatial components $A_k = A^k$ with $k=1,2,3$. Peierls substitution results in the dependence of $Q$ on the combination $\pi = p-A(x)$:  ${Q}_W(p-A(x))=Q_W(\pi)$.
If in addition there is another reason of inhomogeneity, then we can represent
$Q_W(p,x) = Q_W(p-A(x),x)$.

For definiteness we may chose lattice regularization with  Wilson fermions that corresponds to 
\begin{equation}\label{WF}
Q_W(\pi)=\sum_{\mu=1}^3\gamma^{\mu}g_{\mu}(\pi)-ig_5(\pi)+\gamma^4g_4(\pi_4+i \mu_5\gamma^5).
\end{equation}
Here $\gamma^i$ are Euclidean gamma - matrices, while $g_i=\sin(\pi_i)$ with $i=1,2,3,4$;  $g_5(\pi) = m(\pi)=m^{(0)}+\sum\limits_{i=1}^{4}\left(1-\cos(\pi_i)\right)$. $\mu_5$ is chiral chemical potential. For the massless fermions we have  $m^{(0)}=0$. 

We will also consider as a particular example the lattice Dirac operator with the form of 
\begin{equation}\label{WF_}
	Q_W(\pi,x)=\sum_{\mu=1}^3\gamma^{\mu}g_{\mu}(\pi,x)-ig_5(\pi,x)+\gamma^4g_4(\pi_4+i \mu_5\gamma^5,x).
\end{equation}
where functions $g_\mu$ depend on $\pi = p-A(x)$, and in addition depend on $x$ explicitly.

However, the results discussed below remain valid also for the lattice regularization with arbitrary form of $Q_W(p,x)$ provided that the inhomogeneity is negligible at the distance of the order of lattice spacing.

\subsection{Linear response of Green function to external magnetic field}

In the presence of external magnetic field related to Euclidean four - potential $A$ we have
$Q_W(p,x)=Q_W(p-A(x),x)$. Here dependence on magnetic field is encoded in vector potential $A(x)$ while dependence on scalar potential is written as explicit dependence of $Q$ on $x$. The direct dependence of $Q$ on $x$ may also originate from the other sources like the chiral chemical potential depending on $x$ (see below). The Dirac operator may be written as 
$$
Q_W(p,x)\approx Q_W^{(0)}(p,x)+\delta Q_W(p,x)
$$
where $\delta Q_W=-{\partial_p}_kQ^{(0)}(p,x)A_k$ while $Q^{(0)}$ is Wilson Dirac operator with $A=0$. The Green's function may also be represented as 
$$
G_W(p,x)\approx G_W^{(0)}(p,x)+\delta G_W(p,x).
$$
The Groenewold equation results in 
\begin{eqnarray}
&&\left(Q_W^{(0)}(p,x)+\delta Q_W(p,x)\right)\star\left(G_W^{(0)}(p,x)+\delta G_W(p,x)\right)\\\nonumber&=& Q^{(0)}_W\star G^{(0)}_W+Q^{(0)}_W\star \delta G_W+\delta Q_W\star G^{(0)}_W=1,
\end{eqnarray}
where $Q^{(0)}_W\star G^{(0)}_W=1$. We denote the field strength by  $F_{ij}=\partial_iA_j-\partial_jA_i$. Hence we can write $\delta G_W$ as
\begin{eqnarray}
\delta G_W&=&G^{(1)}_{W(k)}A_k+G^{(2)}_{W(ij)}F_{ij}\nonumber\\&=&\left[G^{(0)}_W\star{\partial_p}_kQ^{(0)}_W(p,x)\star G^{(0)}_W\right]A_k+\frac{i}{2}\left[G^{(0)}_W\star{\partial_p}_iQ^{(0)}_W(p,x)\star G^{(0)}_W\star{\partial_p}_jQ^{(0)}_W(p,x)\star G^{(0)}_W\right]F_{ij}\nonumber.
\end{eqnarray}
So, we have 
$$
G_W(p,x) \approx G^{(0)}_W(p,x)+G^{(1)}_{W(k)}A_k+G^{(2)}_{W(ij)}F_{ij}
$$
where \cite{ZS2020}, 
$$
G^{(1)}_{W(k)}=G^{(0)}_W\star{\partial_p}_kQ^{(0)}_W(p,x)\star G^{(0)}_W
$$
$$
G^{(2)}_{W(ij)}=\frac{\ii}{2}\left[G^{(0)}_W\star{\partial_p}_iQ^{(0)}_W(p,x)\star G^{(0)}_W\star{\partial_p}_jQ^{(0)}_W(p,x)\star G^{(0)}_W\right]
$$
\subsection{Electric current}

In the presence of gauge field electric current may be obtained varying the partition function 
$$
j_k(x)=\frac{\delta \rm ln Z}{\delta A_k(x)}=-\frac{1}{2}\int_{\mathcal M}\frac{d^Dp}{|\mathcal M|}\rm tr\left[G_W(p,x) {\partial_p}_kQ_W(p,x)\right]+c.c
$$
Integral here is over momentum space $\cal M$. Its volume is equal to ${\cal M}=(2 \pi)^D$ for the model with Wilson fermions. The total integrated current is equal to the integral over coordinates
$$
J_k=\int dx j_k(x)=-\frac{1}{2} \int d^{D}x\int_{\mathcal M}\frac{d^Dp}{|\mathcal M|}\rm tr\left[G_W(p,x) {\partial_p}_kQ_W(p,x)\right]+c.c.
$$
Notice, that here and below we replace in such expressions sum over the lattice points by an integral, which is possible since we are considering the model with the inhomogeneity that is negligible at the distance of the lattice spacing.
Using periodic boundary condition, we have 
\begin{eqnarray}
J_k&=&-\int d^{D}x\int_{\mathcal M}\frac{d^Dp}{|\mathcal M|}\rm \left[G_W(p,x){\partial_p}_kQ_W(p,x)\right]\nonumber\\&=&-\int d^{D}x\int_{\mathcal M}\frac{d^Dp}{|\mathcal M|}\rm \left[G_W(p,x)\star{\partial_p}_kQ_W(p,x)\right].
\end{eqnarray}
Here ${\partial_p}_kQ_W(p,x)={\partial_p}_kQ^{(0)}_W(p,x)-\left({\partial_p}_k{\partial_p}_jQ^{(0)}_W(p,x)\right)A_j.$ So we have
$$
G_W(x,p){\partial_p}_kQ_W(p,x)=\left(G^{(0)}_W(p,x)+G^{(1)}_{W(l)}A_l+G^{(2)}_{W(mn)}F_{mn}\right)\left({\partial_p}_kQ^{(0)}_W(p,x)-\left({\partial_p}_i
{\partial_p}_jQ^{(0)}_W(p,x)\right)A_j\right).
$$
Current density is given by 
\begin{eqnarray}
j_i(x)&=&-\int_{\mathcal M}\frac{d^Dp}{|\mathcal M|}\rm Tr\Large[G^{(0)}_W{\partial_p}_iQ^{(0)}_W-{\partial_p}_k\left(G^{(0)}_W{\partial_p}_iQ^{(0)}_W\right)A_k\nonumber\\&-&\frac{i}{2}\left[G^{(0)}_W\star{\partial_p}_mQ^{(0)}_W(p,x)\star G^{(0)}_W\star{\partial_p}_nQ^{(0)}_W(p,x)\star G^{(0)}_W\right]{\partial_p}_iQ^{(0)}_W F_{mn}\Large].
\end{eqnarray}
We denote 
$$
j_i(x)=j_i^{(0)}(x)+j_{i(k)}^{(1)}(x)A_k(x)+j_{i(mn)}^{(2)}(x)F_{mn}(x)
$$
where
$$
j_i^{(0)}(x)=-\int_{\mathcal M}\frac{d^Dp}{|\mathcal M|}\rm Tr\left[G^{(0)}_W{\partial_p}_iQ^{(0)}_W\right] ~~\text{and}~~
j_{i(k)}^{(1)}(x)=\int_{\mathcal M}\frac{d^Dp}{|\mathcal M|}\rm Tr
\left[{\partial_p}_k\left(G^{(0)}_W{\partial_p}_iQ^{(0)}_W\right)\right].$$
For the periodic boundary conditions in momentum space $j_{i(k)}^{(1)}(x)=0$ since it is an integral of total derivative. The first leading order term with the external gauge field is given by
$$
j_{i(mn)}^{(2)}(x)=-\frac{i}{2}\int_{\mathcal M}\frac{d^Dp}{|\mathcal M|}\rm Tr\left[\left[G^{(0)}_W\star{\partial_p}_mQ^{(0)}_W(p,x)\star G^{(0)}_W\star{\partial_p}_nQ^{(0)}_W(p,x)\star G^{(0)}_W\right]{\partial_p}_iQ^{(0)}_W\right].
$$
The average current linear in $F_{mn}$ is
\begin{eqnarray}
\bar{J}_i^{(2)}&=&\frac{J_i^{(2)}}{V^{(4)}}\equiv \frac{F_{mn}}{V^{(4)}}\int d^Dx j^{(2)}_{i(mn)}(x)=\frac{T F_{mn}}{V}\int dx j^{(2)}_{i(mn)}(x)\\\nonumber&=&\frac{-iF_{mn}}{2V}\int d^Dx \int_{\mathcal M}\frac{d^Dp}{|\mathcal M|}\rm Tr\left[\left[G^{(0)}_W\star{\partial_p}_mQ^{(0)}_W(p,x)\star G^{(0)}_W\star{\partial_p}_nQ^{(0)}_W(p,x)\star G^{(0)}_W\right]{\partial_p}_iQ^{(0)}_W\right],
\end{eqnarray}
Here $V^{(4)}$ is the overall volume of Euclidean space-time while $V$ is the three-dimensional volume, $T$ is temperature assumed to be very small. 
We have the average current  \cite{ZW2019}
$$
\bar{J}^k=\frac{1}{4\pi^2}\epsilon^{ijkl}\mathcal{M}_lF_{ij}
$$
where
$$
\mathcal{M}_l=\frac{-iT\epsilon_{ijkl}}{3!V8\pi^2}\int d^Dx \int_{\mathcal M}{d^Dp}\rm Tr\left[\left[G^{(0)}_W\star{\partial_p}_iQ^{(0)}_W(p,x)\star G^{(0)}_W\star{\partial_p}_jQ^{(0)}_W(p,x)\star G^{(0)}_W\right]{\partial_p}_kQ^{(0)}_W\right]
$$
In the last expression the star may be inserted and we will arrive at 
$$
\mathcal{M}_l=\frac{-iT\epsilon_{ijkl}}{3!V8\pi^2}\int d^Dx \int_{\mathcal M}{d^Dp}\rm Tr\left[G^{(0)}_W\star{\partial_p}_iQ^{(0)}_W(p,x)\star G^{(0)}_W\star{\partial_p}_jQ^{(0)}_W(p,x)\star G^{(0)}_W\star {\partial_p}_kQ^{(0)}_W\right]
$$
This expression is topological invariant as long as the system is gapped and the integral is convergent.
We consider the magnetic field only, which is calculated through the field tensor $F_{ij}=\epsilon_{ijk}B_k$. Notice, that we define the field strength in Euclidean space - time, and assume that $A^k = A_k$. Therefore, from ${\bf B} = {\rm rot}\, {\bf A}$ it follows $B_k = \epsilon_{kij} \partial_i A^j$. In this case and in the presence of finite chiral chemical potential the system remains gapped in the absence of inhomogeneity. This property is kept also when weak inhomogeneity is added. Then the topological nature of quantity $M_4$ allows us to come to the conclusion that the CME conductivity vanishes also in the presence of weak inhomogeneity. This is an extension of the result of \cite{Z2016} \footnote{Notice, that in \cite{Z2016} there was a mistake (wrong sign) in the second row of Eq. (8), consequently the signs are to be changed to inverse in Eqs. (10), (11), (32), (33), (34), (35), (38), the second and the third rows in Eq. (31) of \cite{Z2016};   see also corrigendum in \cite{ZW2019}. Fortunately, this wrong sign did not affect the physical results of \cite{Z2016}. }.

\section{Theory at nonzero temperature}

\subsection{Chiral magnetic conductivity}

At finite temperature 
the partition function receives the form
\begin{eqnarray}
	Z &=& \int D\bar{\psi}D\psi \, {\rm exp}\Big( 2\pi T  \sum_{p_1^4=p_2^4=\omega}\int \frac{d^{D-1} {p}_1}{\sqrt{|{\cal M}|}} \int \frac{d^{D-1} {p}_2}{\sqrt{|{\cal M}|}}\bar{\psi}^T({p}_1){\cal Q}(p_1,p_2)\psi({p}_2) \Big),\label{Z1}
\end{eqnarray}
Here $\sum_{\omega}$ is the sum over Matsubara frequencies $\omega = 2 \pi T (n+1/2)$, $n \in Z$, $0\le n < N$, where $N = 1/T$ is the number of lattice sites in the fourth direction, while temperature $T$ is measured in the lattice units. Integrals are over the Brillouin zone $\cal B$. Notice that the volume of momentum space at zero temperature is equal to $2\pi$ times $|{\cal B}|$, where $|{\cal B}|$ is volume of the three - dimensional Brillouin zone.

In the similar way we obtain that the chiral magnetic current averaged over the sample volume is given by 
$$
\bar{J}^k=\frac{1}{4\pi^2}\epsilon^{ijk4}\mathcal{M}_4F_{ij}.
$$
Here  
$$
\mathcal{M}_4=\frac{-2\pi i T\epsilon_{ijk4}}{3!V8\pi^2}\int d^{D-1} x \int_{\mathcal B}{d^{D-1}p} \sum\limits_{\omega}\rm Tr\left[\left[G^{(0)}_W\star{\partial_p}_iQ^{(0)}_W(p,x)\star G^{(0)}_W\star{\partial_p}_jQ^{(0)}_W(p,x)\star G^{(0)}_W\right]{\partial_p}_kQ^{(0)}_W\right].
$$
and $\sum_{\omega}$ is the sum over Matsubara frequencies $\omega = 2 \pi T (n+1/2)$, $n \in Z$, $0\le n < N$, where $N = 1/T$.
We are able to add an extra star to the above expression and arrive at
$$
\mathcal{M}_4 = 2\pi T \sum\limits_{\omega} \mathcal{N}_4(\omega)
$$
where
\begin{equation}
\mathcal{N}_4(\omega)=\frac{- i \epsilon_{ijk4}}{3!V8\pi^2}\int d^{D-1} x \int_{\mathcal B}{d^{D-1}p} \rm Tr\left[G^{(0)}_W\star{\partial_p}_iQ^{(0)}_W(p,x)\star G^{(0)}_W\star{\partial_p}_jQ^{(0)}_W(p,x)\star G^{(0)}_W\star{\partial_p}_kQ^{(0)}_W\right].
\end{equation}
Since $Q_W$ and $G_W$ do not depend explicitly on imaginary time $\tau$, the star operation is reduced to the one without derivatives with respect to $\omega$ and $\tau$. One can see that for any given value of $\omega$ quantity $\mathcal{N}_4(\omega)$ is a topological invariant as long as the expression standing inside the integral does not contain  singularity (the proof is similar to that of the similar statement presented in \cite{ZW2019}). We know that finite temperature being added to the homogeneous model of massless fermions removes the singularity of the Green function. Weak inhomogeneity cannot change this. We come to conclusion that the response of $\mathcal{M}_4$ to chiral chemical potential at finite temperature (i.e. the CME conductivity) vanishes even for the system with inhomogeneity, at least if the inhomogeneity is sufficiently weak. 

Below we will check by direct calculation that the CME conductivity indeed vanishes for the particular case of the model with the Dirac operator given by Eq. (\ref{WF_}). An example of such a model is given by Wilson fermions, or by more complicated but still standard lattice regularizations. 

We will denote $r = \frac{2\pi T}{16 V \pi^2}$ and consider the linear response of $\mathcal{M}_4$ to $\mu_5$: 
\begin{equation}
\mathcal{M}_4=\mathcal{M}_4|_{\mu_5=0}+\mu_5\frac{\partial \mathcal{M}_4}{\partial\mu_5}
\end{equation}
We represent 
\begin{eqnarray}
\frac{\partial \mathcal{M}_4}{\partial\mu_5}&=&r\,  \epsilon_{ijk}\int d^3x \int_{\mathcal B}{d^3p} \sum\limits_{\omega}\rm Tr\LARGE[\gamma^5G^{(0)}_W\star{\partial_p}_iQ^{(0)}_W\star G^{(0)}_W\\\nonumber&&\star{\partial_p}_jQ^{(0)}_W\star G^{(0)}_W\star{\partial_p}_kQ^{(0)}_W\star G^{(0)}_W\star{\partial_p}_4Q^{(0)}_W\LARGE]\\\nonumber&=&r\, \epsilon_{ijk}\int d^3x \int_{\mathcal B} {d^3p} \sum\limits_{\omega}\rm Tr\left[\gamma^5{\partial_p}_iG^{(0)}_W\star{\partial_p}_jQ^{(0)}_W\star {\partial_p}_kG^{(0)}_W\star {\partial_p}_4Q^{(0)}_W\right].\label{sigmaCME}
\end{eqnarray}

\subsection{Homogeneous limit}

For simplicity first let us consider the limit of homogeneous system without  explicit dependence of $Q$ on spatial coordinates.  
We represent $Q=\sum_{\mu} g_{\mu}\gamma^{\mu}-i g_5=-i \gamma^5\sum_{a=1,2,3,4,5}\Gamma^ag_a=-i \gamma^5\tilde{Q}$ where $\tilde{Q}=\sum_a\Gamma^ag_a$. Here $\Gamma^5=\gamma^5$ and $\Gamma^{\mu}=i\gamma^5\gamma^{\mu}$ for $\mu=1,2,3,4$ and it satisfies the usual anti-commutation relations of Dirac gamma matrices; $\{\Gamma^a,\Gamma^b\}=2\delta^{ab}$. So we have $G=\frac{\sum_a\Gamma^ag_a}{||g||^2}\times (i \gamma^5)=\tilde{G}\times (i \gamma^5)$ where $\tilde{G}=\sum_a\Gamma^af_a$. We hereafter replace the star product by ordinary product and write the above equation in terms of $\tilde{Q}$ and $\tilde{G}$ which is given by
\begin{eqnarray}
\frac{\partial \mathcal{M}_4}{\partial\mu_5}&=&r\, \epsilon_{ijk}\int d^3x \int_{\mathcal B}{d^3p} \sum\limits_{\omega}\rm Tr\left[\gamma^5{\partial_p}_iG^{(0)}_W{\partial_p}_jQ^{(0)}_W {\partial_p}_kG^{(0)}_W {\partial_p}_4Q^{(0)}_W\right]\\\nonumber&=&r\, \epsilon_{ijk}\int d^3x \int_{\mathcal B}{d^3p}\sum\limits_{\omega}\rm Tr\left[\gamma^5{\partial_p}_i\tilde{G}^{(0)}_W(i\gamma^5)(-i\gamma^5){\partial_p}_j\tilde{Q}^{(0)}_W {\partial_p}_k\tilde{G}^{(0)}_W(i\gamma^5)(-i\gamma^5) {\partial_p}_4\tilde{Q}^{(0)}_W\right]\\\nonumber&=&r\, \epsilon_{ijk}\int d^3x \int_{\mathcal B}{d^3p}\sum\limits_{\omega}\rm Tr\left[\gamma^5{\partial_p}_i\tilde{G}^{(0)}_W{\partial_p}_j\tilde{Q}^{(0)}_W {\partial_p}_k\tilde{G}^{(0)}_W {\partial_p}_4\tilde{Q}^{(0)}_W\right].
\end{eqnarray}
For $\tilde{Q}=\sum_a\Gamma^ag_a$ and $\tilde{G}=\frac{\sum_a\Gamma^ag_a}{||g||^2}$, we have 
\begin{eqnarray}
\frac{\partial \mathcal{M}_4}{\partial\mu_5}=r\, \epsilon_{ijk}\int d^3x \int_{\mathcal B}{d^3p} \sum\limits_{\omega}\rm Tr\left[\Gamma^5\Gamma^a\Gamma^b\Gamma^c\Gamma^d\right]{\partial_p}_i\left(\frac{g_a}{||g||^2}\right){\partial_p}_jg_b{\partial_p}_k\left(\frac{g_c}{||g||^2}\right){\partial_p}_4g_d.
\end{eqnarray}

Using results of Appendix \ref{SectTr} we come to the following expression of $\frac{\partial \mathcal{M}_4}{\partial\mu_5}$:
\begin{eqnarray}
\frac{\partial \mathcal{M}_4}{\partial\mu_5}=4\, r\,\epsilon_{ijk}\epsilon_{\mu\nu\rho\sigma}\int d^3x \int_{\mathcal B}{d^3p}\sum\limits_{\omega}{\partial_p}_i\left(\frac{g_{\mu}}{||g||^2}\right){\partial_p}_jg_{\nu}{\partial_p}_k\left(\frac{g_{\rho}}{||g||^2}\right){\partial_p}_4g_{\sigma}.
\end{eqnarray}
Let us represent 
\begin{eqnarray}
{\partial_p}_i\left( \frac{g_{\mu}}{||g||^2}{\partial_p}_jg_{\nu}{\partial_p}_k\left(\frac{g_{\rho}}{||g||^2}\right){\partial_p}_4g_{\sigma}\right)&=&{\partial_p}_i\left(\frac{g_{\mu}}{||g||^2}\right){\partial_p}_jg_{\nu}{\partial_p}_k\left(\frac{g_{\rho}}{||g||^2}\right){\partial_p}_4g_{\sigma}\\\nonumber&+&\frac{g_{\mu}}{||g||^2}{\partial_p}_i{\partial_p}_jg_{\nu}{\partial_p}_k\left(\frac{g_{\rho}}{||g||^2}\right){\partial_p}_4g_{\sigma}
\\\nonumber&+&\frac{g_{\mu}}{||g||^2}{\partial_p}_jg_{\nu}{\partial_p}_i{\partial_p}_k\left(\frac{g_{\rho}}{||g||^2}\right){\partial_p}_4g_{\sigma}
\\\nonumber&+&\frac{g_{\mu}}{||g||^2}{\partial_p}_jg_{\nu}{\partial_p}_k\left(\frac{g_{\rho}}{||g||^2}\right){\partial_p}_i{\partial_p}_4g_{\sigma}.
\end{eqnarray}
Here ${\partial_p}_4g_{\sigma}=f(\omega)$ and ${\partial_p}_i{\partial_p}_4g_{\sigma}=0$ and 
$\epsilon_{ijk}\frac{g_{\mu}}{||g||^2}{\partial_p}_i{\partial_p}_jg_{\nu}{\partial_p}_k\left(\frac{g_{\rho}}{||g||^2}\right){\partial_p}_4g_{\sigma}=0$ and $\epsilon_{ijk}\frac{g_{\mu}}{||g||^2}{\partial_p}_jg_{\nu}{\partial_p}_i{\partial_p}_k\left(\frac{g_{\rho}}{||g||^2}\right){\partial_p}_4g_{\sigma}=0.$
We come to the expression for the CME conductivity equal to the integral of total derivative 
\begin{eqnarray}
\frac{\partial \mathcal{M}_4}{\partial\mu_5}&=&4\, r\,\epsilon_{ijk}\epsilon_{\mu\nu\rho\sigma}\int d^3x \int_{\mathcal B}{d^3p}\sum\limits_{\omega}{\partial_p}_i\left( \frac{g_{\mu}}{||g||^2}{\partial_p}_jg_{\nu}{\partial_p}_k\left(\frac{g_{\rho}}{||g||^2}\right){\partial_p}_4g_{\sigma}\right)
\\\nonumber&=&0.
\end{eqnarray}

\subsection{Calculation for the non - homogeneous case}

Now let us extend the above calculation to the non - homogeneous case. We still may represent $$Q_W=\sum_{\mu} g_{\mu}\gamma^{\mu}-i g_5=-i \gamma^5\sum_{a=1,2,3,4,5}\Gamma^ag_a=-i \gamma^5\tilde{Q}$$
 where $\tilde{Q}=\sum_a\Gamma^ag_a$. This is the same expression as for the homogeneous case, except that functions $g_\mu$ depend both on momenta and on spatial coordinates. Here $\Gamma^5=\gamma^5$ and $\Gamma^{\mu}=i\gamma^5\gamma^{\mu}$ for $\mu=1,2,3,4$ and it satisfies the usual anti-commutation relations of Dirac gamma matrices; $\{\Gamma^a,\Gamma^b\}=2\delta^{ab}$. For the Wigner transformation of Green function we have a more complicated expression
  $$
 G_W=(f+\sum_a\Gamma^a f_a + \sum_{ab} f_{ab}\Gamma^a \Gamma^b)\times (i \gamma^5)=\tilde{G}\times (i \gamma^5)
 $$
  where $$\tilde{G}_W=f+\sum_a\Gamma^a f_a + \sum_{ab} f_{ab}\Gamma^a \Gamma^b$$
In order to determine functions $f(p,x), f_a(p,x), f_{ab}(p,x)$ we substitute the above expression to the Groenewold equation. In particular, this allows us to obtain $f=0$, and we come to the following expression for $\frac{\partial \mathcal{M}_4}{\partial\mu_5}$:
\begin{eqnarray}
	\frac{\partial \mathcal{M}_4}{\partial\mu_5}&=& 4\, r\,\epsilon_{ijk}\epsilon_{\mu\nu\rho\sigma}\int d^3x \int_{\mathcal B}{d^3p}\sum\limits_{\omega}{\partial_p}_i f_\mu\star{\partial_p}_jg_{\nu}\star{\partial_p}_k  f_\rho\star {\partial_p}_4g_{\sigma}\nonumber\\
	&&+ 4\, r\,\epsilon_{ijk}{\cal Y}_{\mu  \nu \rho_1 \rho_2 \sigma}\int d^3x \int_{\mathcal B}{d^3p}\sum\limits_{\omega}{\partial_p}_i f_{\mu} \star{\partial_p}_jg_{\nu}\star{\partial_p}_k  f_{\rho_1 \rho_2}\star {\partial_p}_4g_{\sigma}\nonumber\\
	&&+ 4\, r\,\epsilon_{ijk}{\cal Y}_{\mu_1 \mu_2 \nu\rho\sigma}\int d^3x \int_{\mathcal B}{d^3p}\sum\limits_{\omega}{\partial_p}_i f_{\mu_1 \mu_2} \star{\partial_p}_jg_{\nu}\star{\partial_p}_k  f_\rho\star {\partial_p}_4g_{\sigma}\nonumber\\
	&&+ 4\, r\,\epsilon_{ijk}{\cal Y}_{\mu_1 \mu_2 \nu\rho_1 \rho_2\sigma}\int d^3x \int_{\mathcal B}{d^3p}\sum\limits_{\omega}{\partial_p}_i f_{\mu_1 \mu_2} \star{\partial_p}_jg_{\nu}\star{\partial_p}_k  f_{\rho_1 \rho_2}\star {\partial_p}_4g_{\sigma}.
\end{eqnarray}
Here we denote
$$
{\cal Y}_{\mu...\sigma} = \frac{1}{4} {\rm tr} \, \Gamma^5 \Gamma^\mu ... \Gamma^\sigma
$$
As above we obtain
  \begin{eqnarray}
  	\frac{\partial \mathcal{M}_4}{\partial\mu_5}&=& 4\, r\,\epsilon_{ijk}\epsilon_{\mu\nu\rho\sigma}\int d^3x \int_{\mathcal B}{d^3p}\sum\limits_{\omega}{\partial_p}_i\Big( f_\mu\star{\partial_p}_jg_{\nu}\star{\partial_p}_k  f_\rho\star {\partial_p}_4g_{\sigma}\Big)\nonumber\\
  	&&+ 4\, r\,\epsilon_{ijk}{\cal Y}_{\mu  \nu \rho_1 \rho_2 \sigma}\int d^3x \int_{\mathcal B}{d^3p}\sum\limits_{\omega}{\partial_p}_i\Big( f_{\mu} \star{\partial_p}_jg_{\nu}\star{\partial_p}_k  f_{\rho_1 \rho_2}\star {\partial_p}_4g_{\sigma}\Big)\nonumber\\
  	&&+ 4\, r\,\epsilon_{ijk}{\cal Y}_{\mu_1 \mu_2 \nu\rho\sigma}\int d^3x \int_{\mathcal B}{d^3p}\sum\limits_{\omega}{\partial_p}_i \Big(f_{\mu_1 \mu_2} \star{\partial_p}_jg_{\nu}\star{\partial_p}_k  f_\rho\star {\partial_p}_4g_{\sigma}\Big)\nonumber\\
  	&&+ 4\, r\,\epsilon_{ijk}{\cal Y}_{\mu_1 \mu_2 \nu\rho_1 \rho_2\sigma}\int d^3x \int_{\mathcal B}{d^3p}\sum\limits_{\omega}{\partial_p}_i\Big( f_{\mu_1 \mu_2} \star{\partial_p}_jg_{\nu}\star{\partial_p}_k  f_{\rho_1 \rho_2}\star {\partial_p}_4g_{\sigma}\Big).
  \end{eqnarray}
  This is again the  integral of total derivative, and therefore is equal to zero.
  
  \section{Effect of interactions}
  
  \begin{figure}[h]
  	\centering  %
  	\includegraphics[height=5cm]{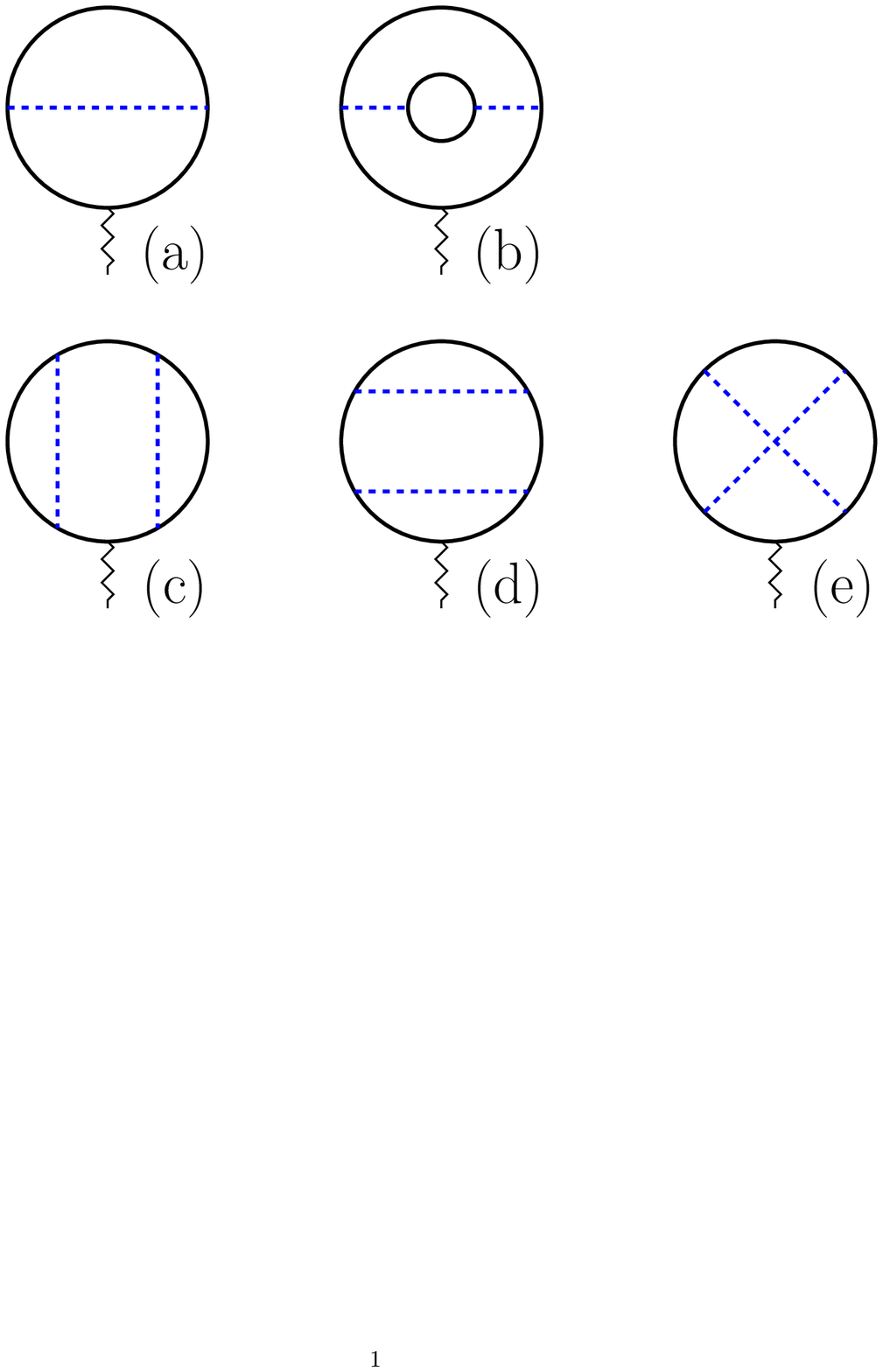} \vspace{1cm} %
  	\caption{The diagrams for fermion propagator with one and two loops, and their  contributions to electric current. }  %
  	\label{fig.5}   %
  \end{figure}

Let us consider briefly effect of interactions on the CME conductivity.  Without interactions the averaged electric current in the presence of external magnetic field is given by    
  \begin{eqnarray}
  	\bar{J}_k&=&-2\pi T\int \frac{{d^{D-1}x}}{V}\sum_{p_4} \int_{\mathcal B}\frac{d^{D-1}p}{|\mathcal M|}{\rm Tr} \rm \left[G_W(x,p)\star{\partial_p}_kQ_W(p,x)\right].\label{Jk}
  \end{eqnarray} 
Here $G_W$ is Wigner transformation of fermion propagator in the presence of magnetic field, while $Q_W$ is Weyl symbol of its inverse operator. For definiteness we consider interaction caused by gauge bosons. Then in one loop approximation (Fig. \ref{fig.5} a)) we obtain the following correction to electric current:
\begin{eqnarray}
	\bar{J}^{(1)}_k&=&-(2\pi T)^2\int \frac{d^{D-1}x}{V}\sum_{p_4,k_4}\int_{\mathcal B}\frac{d^{D-1}p}{|\mathcal M|}{\rm Tr}\rm \left[\gamma^\mu \hat{t}^a G_W(x,p-k)\gamma^\nu \hat{t}^b \star{\partial_p}_kG_W(p,x)\right]\frac{d^{D-1} k}{|\mathcal M|}D^{ab}_{\mu \nu}(k)
\end{eqnarray}   
 Here $D^{ab}_{\mu \nu}(k)$ is propagator of gauge boson while $\hat{t}^a$ is generator of the gauge group. We assume that bare action of the gauge boson is unform, i.e. bare Wigner transformed propagator depends on momentum only and does not depend on coordinates.  Using integration by parts we represent 
 \begin{eqnarray}
 	\bar{J}^{(1)}_k&=&-(2\pi T)^2\int \frac{d^{D-1}x}{V}\sum_{p_4, k_4}\int_{\mathcal B}\frac{d^{D-1}p}{|\mathcal M|}{\rm Tr}\rm \left[\gamma^\mu \hat{t}^a G_W(x,p-k)\gamma^\nu \hat{t}^b \star{\partial_p}_kG_W(p,x)\right]\frac{d^{D-1} k}{|\mathcal M|}D^{ab}_{\mu \nu}(k)\nonumber\\
 	&=&(2\pi T)^2\int \frac{d^{D-1}x}{V}\sum_{p_4,k_4}\int_{\mathcal B}\frac{d^{D-1}p}{|\mathcal M|}{\rm Tr}\rm \left[\gamma^\mu \hat{t}^a {\partial_p}_k G_W(x,p-k)\gamma^\nu \hat{t}^b \star G_W(p,x)\right]\frac{d^{D-1} k}{|\mathcal M|}D^{ab}_{\mu \nu}(k) \nonumber\\
 	&=&(2\pi T)^2\int \frac{d^{D-1}x}{V}\sum_{p_4,k_4}\int_{\mathcal B}\frac{d^{D-1}p}{|\mathcal M|}{\rm Tr}\rm \left[\gamma^\mu \hat{t}^a G_W(x,p-k)\gamma^\nu \hat{t}^b \star{\partial_p}_kG_W(p,x)\right]\frac{d^{D-1} k}{|\mathcal M|}D^{ba}_{\nu \mu}(-k)\label{oneloop}
 \end{eqnarray} 
We can always chose the gauge, in which the gauge boson propagator obeys
 $$
 D^{ba}_{\nu \mu}(-k) = D^{ab}_{\mu \nu}(k)
 $$ 
  As a result we obtain
  $$
  \bar{J}^{(1)}_k = - \bar{J}^{(1)}_k = 0
  $$

\begin{figure}[h]
	\centering  %
	\includegraphics[height=2.5cm]{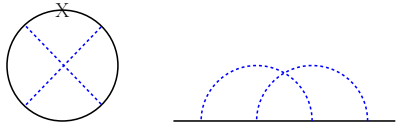}\vspace{1cm}  %
	\caption{An example of the two loop contribution to self energy, which gives rise to a three loop contribution to electric current $J^{(2,cross)}_k$. The cross marks the position of derivative $\partial_{p_k}Q_{W}$.}  %
	\label{fig.2}   %
\end{figure}

For Abelian theory this proof may be extended to all orders of perturbation theory along the lines of \cite{ZZ2019}, which will give that the total integrated electric current is not renormalized by interactions. The same may be true also for the case of the non - Abelian gauge theory, but for the non - Abelian theory we restrict ourselves by the one - loop order. Below we will consider the sketch of the proof that the higer order diagrams do not renormalize electric current for the case of Abelian gauge theory without self - interactions of gauge field quanta. 

Let us consider first the case of the diagram of Fig. \ref{fig.5} b). 
The corresponding contribution to electric current is given by Eq. (\ref{oneloop}) with gauge boson propagator substituted by its one - loop correction
$$
D^{ab}_{\mu \nu}(k) \to D^{(1)ab}_{W,\mu \nu}(x,k)
$$ 
where
$$
D^{(1)ab}_{W,\mu \nu}(x,k) = D^{ac}_{\mu \rho}(k)\star \Pi^{\rho \sigma}_{W,cd}(x,k)\star D^{db}_{\sigma \mu}(k)
$$
Here $\Pi^{\rho \sigma}_{W,cd}(x,k)$ is Wigner transformation of polarization operator  
$$
\Pi^{\rho \sigma}_{cd}(x,y) = {\rm Tr}\gamma^\rho \hat{t}^c G(x,y) \gamma^\sigma \hat{t}^d G(y,x)
$$
One can check that 
$$
D^{(1)ba}_{\nu \mu}(x,-k) = D^{(1)ab}_{\mu \nu}(x,k)
$$
This will give us vanishing contribution to electric current.

The next example is the two - loop diagram of Fig. \ref{fig.5} e). On Fig. \ref{fig.2} we represent the corresponding diagram of self energy. It  gives the contribution to electric current
\begin{eqnarray}\label{current_i2}
	 \bar{J}^{(2,cross)}_k &=& -\frac{(2\pi T)^3}{V|{\cal M}|^3}\sum_{p_4, k_4, q_4} \int { d^{D-1} xd^{D-1} pd^{D-1} k d^{D-1} q} \,{\rm Tr} \Big[\gamma^\mu \hat{t}^a G_{W}(x,p-k)\gamma^\nu \hat{t}^b \star G_{W}(x,p-k-q)\nonumber\\&&\gamma^\rho \hat{t}^c D_{\mu \rho (1)}^{ ac}(k)\star G_{W}(x,p-q)\gamma^\sigma \hat{t}^d\Big]D_{\nu\sigma (2)}^{bd}(q)\partial_{p_k}G_{W}(x,p) \nonumber\\
	&=&-\frac{1}{4}\frac{(2\pi T)^3}{V|{\cal M}|^3}\sum_{p_4, k_4, q_4} \int { d^{D-1} xd^{D-1} pd^{D-1} k d^{D-1} q} \,\partial_{p_k}{\rm Tr} \Big[\gamma^\mu \hat{t}^a G_{W}(x,p-k)\gamma^\nu \hat{t}^b\star G_{W}(x,p-k-q)\nonumber\\&&\gamma^\rho \hat{t}^c D_{\mu \rho (1)}^{ ac}(k)\star G_{W}(x,p-q)\gamma^\sigma \hat{t}^d\Big]D_{\nu\sigma(2)}^{bd}(q)G_{W}(x,p)\nonumber\\&=&0
\end{eqnarray}
\begin{figure}
	\includegraphics[height=3.5cm]{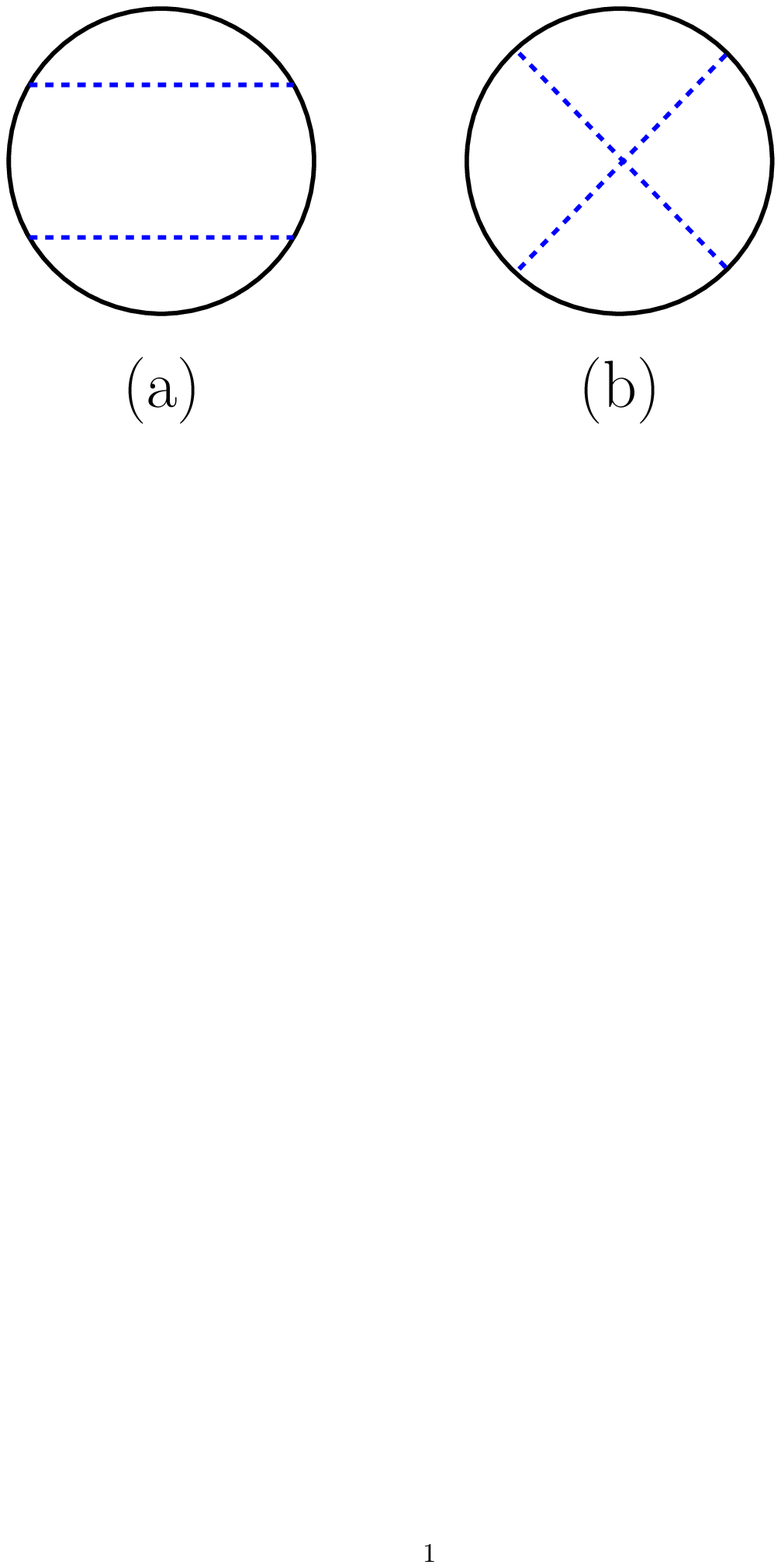}
	\caption{a) The progenitor  for a two - loop rainbow contribution to self energy/electric current. b) The progenitor for the  two - loop cross - contribution to electric current.
	}
	\label{progenitor}
\end{figure}

We use here notations of Feynmann diagram technique designed for the Wigner transformed propagators (it has been proposed in \cite{ZZ2019_}). Here the Moyal product $*=e^{i\overleftarrow{\partial}_x\overrightarrow{\partial}_p/2-i\overleftarrow{\partial}_p\overrightarrow{\partial}_x/2}$ acts on $G$ but does not influence $D$. In \cite{ZZ2019_} also another type of the star product has been introduced  $\circ_{i} = e^{i\overleftarrow{\partial}_R\overrightarrow{\partial}_p/2-i\overleftarrow{\partial}_p\overrightarrow{\partial}_R/2}$. Here derivatives with the right arrow act on Wigner transformed gauge boson propagator $D_{W(i)}(R,k)$ while the derivatives with the left arrow act to all propagators of fermions that stand left to this symbol.  In a similar way in $_i\circ$ the derivatives with the right arrow act on one function that stands right to the symbol, and  the derivatives with the left arrow act on gauge boson propagator $D_{W(i)}(R,k)$. However, at the given order the Wigner transformed boson propagator does not depend on $R$, it depends on momentum ($k$ or $q$) only, and is given by an ordinary Fourier transform. Therefore, the circle products may be omitted completely at this orider. However, they should be present in the next orders of perturbation expansion because the (Wigner transformed) gauge boson propagator becomes depending both on $R$ and $k$ due to the non - homogeneity in $G$.   

\begin{figure}[h]
 	\centering  %
 	\includegraphics[height=5cm]{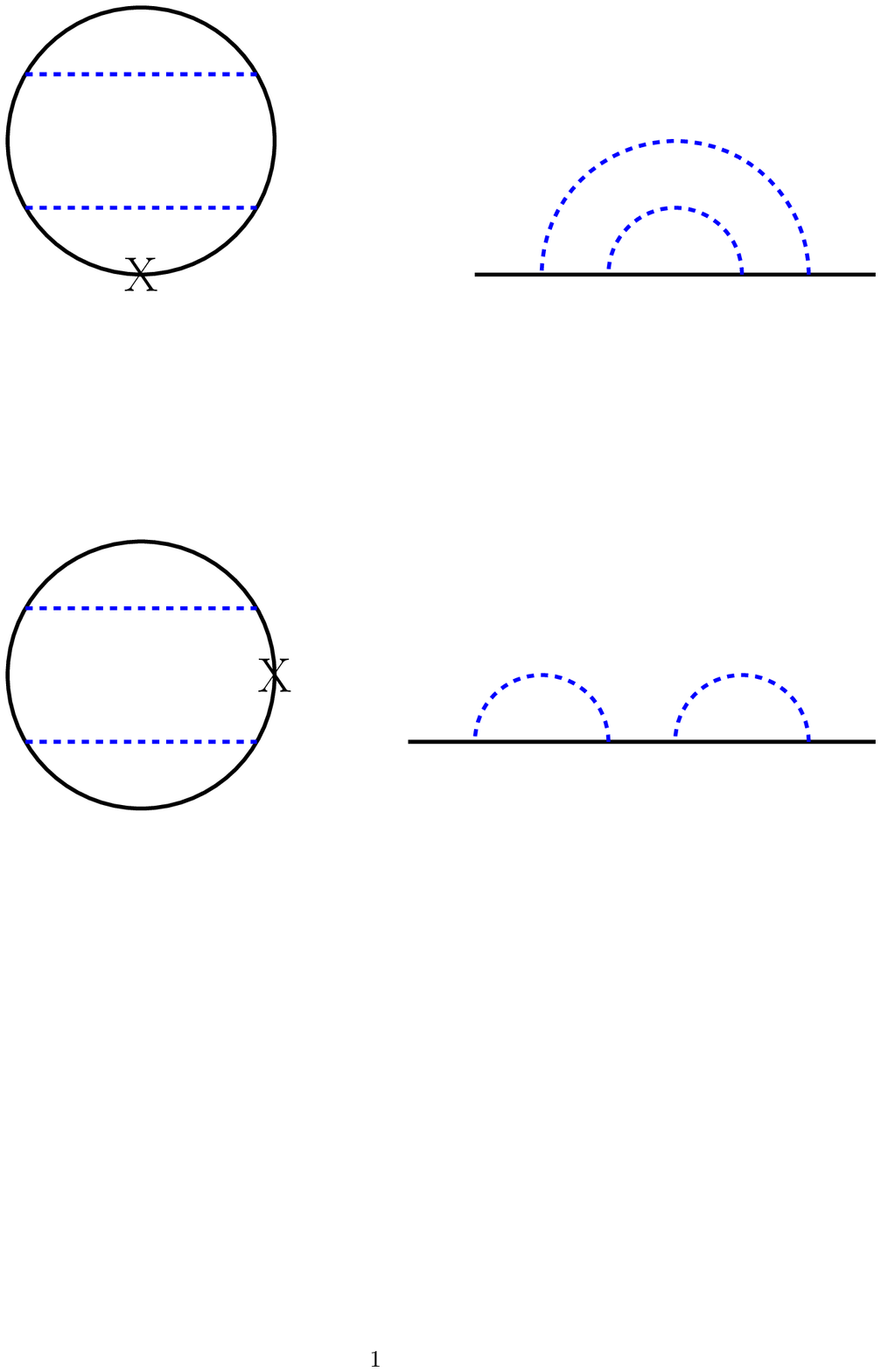} \vspace{1cm} %
 	\caption{Two loop rainbow contributions to self energy, and the corresponding  rainbow diagram for electric current $J^{(2,rainbow)}_k$.  The crosses mark positions of  derivatives $\partial_{p_k}Q_{W}$.}  %
 	\label{fig.1}   %
 \end{figure}

The last expression corresponds to the so - called progenitor diagram of Fig. \ref{progenitor} b) with the derivatives inserted inside the integral over $p$. The integral of a total derivative is zero, and we come to vanishing contribution of the two - loop cross - diagram. Correspondingly, the diagram of Fig. \ref{progenitor} a) is a progenitor for the rainbow contributions to electric current of Fig. \ref{fig.5} c) and d). The corresponence between the two is represented on Fig. \ref{fig.1}. In the similar way it may be shown that the latter contributions vanish as well.

 Moreover, we know that Eq. (\ref{Jk}) is topological invariant. Therefore, in this expression we may replace bare $G_W$ (without interaction corrections) by Wigner transformation of complete interacting Green function and $Q_W$ by Weyl symbol of its inverse. As a result the CME conductivity for the system in the presence of interactions may be written as Eq. (\ref{sigmaCME}) with complete interacting $Q$ and $G$. 
 
\section{Conclusions}

We extended in the present paper the methodology of \cite{Z2016}. The Wigner - Weyl calculus is used to express response of electric current to external magnetic field through the topological invariant composed of the Green functions. In \cite{Z2016} the lattice models of relativistic quantum field theory were discussed. The theory was in thermal equilibrium at zero temperature and was assumed  homogeneous. Besides, the  interactions were not taken into account. Here we take into account dependence of lattice Dirac operator on spatial coordinates. The dependence may be almost arbitrary, but it should be weak in the sense that the variations on the distance of the order of lattice spacing are negligible. In addition we do not restrict ourselves by zero temperature, and consider the theory at arbitrary finite temperature. Finally, we take into account inter - fermion interactions due to exchange by gauge bosons. Explicit calculations are presented in one and two loop approximations, but they may be extended easily to arbitrarily high orders (at least for Abelian gauge group) in direct analogy to the consideration of \cite{ZZ2019}. 

We conclude that the CME conductivity remains vanishing in spite of all considered complications. In particular case of homogeneous system without interactions this conclusion matches the one of \cite{Beneventano:2019qxm}.  It would be interesting to extend the methodology of the present paper to Keldysh technique of kinetic theory in order to find the evidence of the CME in kinetic domain, i.e. either in the presence of both electric and magnetic field \cite{ZrTe5} or without electric field but with the chiral chemical potential depending on time \cite{Wu:2016dam}. 

The authors kindly acknowledge useful discussions with C.X.Zhang. The work of C.Banerjee has been supported by grant of Ministry of Education of Israel.

 \begin{appendix}
 
 \section{Trace calculation}

 \label{SectTr}
 
 Let us calculate the trace of
 $\rm Tr\left[\gamma^5\Gamma^a\Gamma^b\Gamma^c\Gamma^d\right]$ where $a,b,c,d$ run from $1,2,3,4,5$. Here $\Gamma^a=\{\gamma^5,-i\gamma^5\gamma^{\mu}\}$ where $\mu=1,2,3,4$ and it follows the anti-commutation relation $\left\{\Gamma^a,\Gamma^b\right\}=2g^{ab}$.
 When all $\Gamma^{a,b,c,d}=\gamma^5$ then 
 $$
 \rm Tr\left[\gamma^5\Gamma^a\Gamma^b\Gamma^c\Gamma^d\right]=\rm Tr\left[\gamma^5\gamma^5 \gamma^5 \gamma^5 \gamma^5\right]=\rm Tr\left[\gamma^5\right]=0.
 $$
 Now if $\Gamma^a=\gamma^5$, $\Gamma^b=\gamma^5$, $\Gamma^c=\gamma^5$, and $\Gamma^d=-i\gamma^5\gamma^{\mu}$ then 
 $$
 \rm Tr\left[\gamma^5\Gamma^a\Gamma^b\Gamma^c\Gamma^d\right]=\rm Tr\left[\gamma^5\gamma^5\gamma^5\gamma^5(-i\gamma^5\gamma^{\mu})\right]=-i\rm Tr\left[\gamma^5\gamma^{\mu} \right]=0.
 $$
 The above relation holds because $\left\{\gamma^5,\gamma^{\mu}\right\}=0$. Repeating the same steps we can also show if one of the $\Gamma^a$ is $-i\gamma^5\gamma^{\mu}$ and rest of the matrices are $\gamma^5$ then the trace is zero. The another combination is that $\Gamma^a=\gamma^5$, $\Gamma^b=\gamma^5$, $\Gamma^c=-i\gamma^5\gamma^{\mu}$, and $\Gamma^d=-i\gamma^5\gamma^{\nu}$ then
 $$
 \rm Tr\left[\gamma^5\Gamma^a\Gamma^b\Gamma^c\Gamma^d\right]=\rm Tr\left[\gamma^5\gamma^5\gamma^5(-i\gamma^5\gamma^{\mu})(-i\gamma^5\gamma^{\nu})\right]=\rm Tr\left[\gamma^5\gamma^{\mu}\gamma^{\nu}\right]=-g^{\mu\nu}\rm Tr\left[\gamma^5\right]=0.
 $$
 This also holds for other combinations too. Now if three of $\Gamma$ are $-i\gamma^5\gamma^{\mu}$ then 
 $$
 \rm Tr\left[\gamma^5\Gamma^a\Gamma^b\Gamma^c\Gamma^d\right]=\rm Tr\left[\gamma^5\gamma^5(-i\gamma^5\gamma^{\mu})(-i\gamma^5\gamma^{\nu})(-i\gamma^5\gamma^{\rho})\right]=-i\rm Tr\left[\gamma^5\gamma^{\mu}\gamma^{\nu}\gamma^{\rho}\right]=0.
 $$
 Finally we arrive at the point when all the four $\Gamma$ matrices are $-i\gamma^5\gamma^{\mu}$ then we have
 $$
 \rm Tr\left[\gamma^5\Gamma^a\Gamma^b\Gamma^c\Gamma^d\right]=\rm Tr\left[\gamma^5(-i\gamma^5\gamma^{\mu})(-i\gamma^5\gamma^{\nu})(-i\gamma^5\gamma^{\rho})(-i\gamma^5\gamma^{\sigma})\right]=\rm Tr\left[\gamma^5\gamma^{\mu}\gamma^{\nu}\gamma^{\rho}\gamma^{\sigma}\right]=4\epsilon^{\mu\nu\rho\sigma}.
 $$
 Therefore we have $\rm Tr\left[\gamma^5\Gamma^a\Gamma^b\Gamma^c\Gamma^d\right]=\rm Tr\left[\gamma^5\gamma^{\mu}\gamma^{\nu}\gamma^{\rho}\gamma^{\sigma}\right]=4\epsilon^{\mu\nu\rho\sigma}$.
 
 \end{appendix}


\end{document}